\documentstyle[12pt]{article}
\begin{document}
\begin{center}
{\large Antiferromagnetic systems with spin gap: exact results }
\vskip 1cm
                     Indrani Bose\\
                    Department of Physics\\
                    Bose Institute\\
                    93/1, A.P.C.Road\\
                    Calcutta-700 009, India.
\end{center}
\begin{abstract}
We describe some antiferromagnetic systems which exhibit spin gaps. 
We also discuss the effect of doping one such system, namely, the
spin-ladders, with holes. Some model antiferromagnetic systems
with spin gap are reviewed for which exact results are available. 
Exact results for a doped spin-ladder model are also mentioned. 
\end{abstract}

\section*{I. Introduction}
     Recently, several antiferromagnetic (AFM) systems with spin
gap have been discovered. These systems exhibit a variety of
phenomena a full understanding of which is as yet lacking. The
effect of doping on the spin gap, with either hole or magnetic
and non-magnetic impurities, has also been studied. The AFM
Hamiltonian H is generally of the Heisenberg-type with
nearest-neighbour (n.n.) as well as frustrating
further-neighbour interactions included. 
\begin{equation} 
            H \,=\, \sum_{<ij>}\, J_{ij}\, \vec{S_i}\cdot\,\vec{S_j}  
\end{equation}
${\vec S_i}$ is the spin operator at lattice site i and $J_{ij}$ is the
exchange integral for interaction between spins located at sites
i and j. 

     Some theorems and rigorous results are available for the
Hamiltonian (1). The Lieb-Mattis theorem \cite{Lieb} has proved for
general spin and for all dimensions that for a bipartite lattice, 
the entire eigenvalue spectrum satisfies the inequality
\begin{equation}
           E_0\,(S)\, \leq\,E_0\,(S+1)                         
\end{equation}                                                       
where $E_0(S)$ is the minimum energy state corresponding to total spin
S. The weak inequality becomes a strict inequality for a
ferromagnetic (FM) exchange coupling between spins of the same
sublattice. The theorem is valid for any range of exchange
coupling and the proof does not require the existence of
periodic boundary conditions.

     Marshall's sign rule \cite{Marshall} specifies the structure of the
ground state of the spin-$\frac{1}{2}$ AFM on a bipartite lattice 
with n.n. interaction.  The proof can be extended to spin S, 
next-nearest-neighbour
(n.n.n.) FM interaction but not to n.n.n. AFM interaction. The
ground state wave function $\Psi$ is given by
\begin{equation}
    \Psi\,=\,\sum_\mu\,C_\mu\,\mid\,\mu\,\rangle
\end{equation}                  
where$\mid \mu\rangle$ is one of the $2^N$ basis states. According to Marshall's
sign rule, the coefficient $C_{\mu}$   is of the form
\begin{equation}
     C_{\mu}\,=\,(-1)^{p_\mu}\,\,a_\mu
\end{equation}                                                                  
where $a_\mu$   is real and $\geq$ 0 and $ p_\mu$   is the number of up spins on the
A sublattice (the bipartite lattice consists of two sublattices
A and B).

     The Lieb-Mattis  (LSM) theorem\cite{Lieb1}  says that the excitation
spectrum of the S = $\frac{1}{2}$ Heisenberg AFM linear chain with n.n.
interaction is gapless. The proof can be extended to any half-odd
integer spin \cite{Affleck} but not to integer spins giving rise to the famous
`Haldane's conjecture' which we will discuss shortly. 

     The Mermin Wagner theorem \cite{Mermin} proves that there cannot be
any AFM long range order (LRO) at finite temperature T in
dimensions d =1 and 2. Various proofs \cite{Neves,Dyson} exist to show that LRO
exists in the ground state (T = 0) for S $\geq$  1 in d = 2 and
for S $\geq\,\frac{1}{2}$ in d = 3. The proof for the case S = $\frac{1}{2}$, d = 2 is not, however, 
known but approximate calculations have shown the existence of 
LRO in the ground state.

     The Hamiltonian which describes AFMs doped with holes is
the t-J model:
\begin{eqnarray}
H_{t-J}\,&=&\,-\,t\,\sum_{i,j,\sigma}\,(1\,-\,n_{i-\sigma}\,) 
            C_{i,\sigma}^{\dag}\,C_{j,\sigma}\,(1\,-\,n_{j-\sigma})
         +\,J\,\sum_{<ij>}\,\,\vec{S_i}\cdot\,\vec{S_j} 
\end{eqnarray}                                                         
$C^{\dag}$, C are the electron creation and annihilation operators, i, j
denote the lattice sites, $\sigma$ is the spin index and $n_{i,\sigma}$ is the 
occupation number of the ith site with spin $\sigma$. The first term in
the Hamiltonian is the kinetic energy term and operates only in
the subspace of states in which there is no double occupancy. The
last term is the well-known AFM Heisenberg Hamiltonian. The t-J
model is derived from the well-known Hubbard model in the limit
of strong correlation \cite{Gros}. It is this strong correlation which prohibits
the double occupancy of a site. In the half-filled limit, when
there is one electron at each site, the first term of H  is
ineffective and the Hamiltonian reduces to the AFM Heisenberg Hamiltonian. 
The t-J model has been extensively used to study the properties
of doped high-$T_c$ cuprate systems \cite{Dagotto}. In this review, we
describe some AFM systems with spin gap and also a doped AFM
system, namely, the spin-ladder. In Section II, a general
description of AFM systems with spin gap is given. Some examples
of doped AFMs are also given. In Sections III and IV, some AFM models
are described for which exact results are available. Section IV contains
exact results on the hole dynamics in a spin-ladder model. The
review of exact results is by no means exhaustive and describes
mainly the Author's own work.

\section*{II. Antiferromagnetic systems with spin gap}

     In this Section, we describe real AFM systems with spin
gap. The LSM theorem, as we have already mentioned, is applicable
only to half-odd integer spins. Haldane in 1983 \cite{Haldane} made the
conjecture that integer spin chains have a gap in the excitation
spectrum. Experimental realizations of the Haldane-gap systems include
the compounds $ Ni (C_2H_8N_2)_2 NO_2ClO_4$ (NENP),
$Ni(C_3H_{10}N_2)_2NO_2ClO_4$  (NINO) and more recently
$Y_2BaNiO_5$ \cite{Batlogg}. In 1993, Hase et al \cite{Hase} showed that the S =1/2
AFM Heisenberg spin chain compound $CuGeO_3$ undergoes a
spin-Peierls (SP) transition at a temperature $T_{SP} \approx$ 14 K. In
the SP phase, the ground state becomes dimerized in which
successive pairs of sites are brought close together. The spins
in a pair mainly interact with each other forming singlets (S
=0). There is thus an alternation of exchange interaction
strengths $J_1$ and $J_2$. The ground state is non-magnetic and a finite
energy gap exists in the S=1 spin excitation spectrum. $CuGeO_3$ is
the first example of an inorganic system exhibiting the SP transition. 

     Next, we turn to the discussion of spin-ladders
\cite{Dagotto1}. The spins have magnitude 1/2. The simplest spin-ladder
consists of two chains coupled by rungs and interpolates between
1d and 2d AFMs. The Hamiltonian is given by
\begin{eqnarray}
  H_L\,=\,J_{\parallel}\,\sum_{chains}\,\,\vec{S_i}\cdot\,\vec{S_j}                                     
          +\,J_{\perp}\,\sum_{rungs}\,\vec{S_i}\cdot\,\vec{S_j}
\end{eqnarray}                
where $J_{\parallel}$ and $J_{\perp}$ are the exchange interactions along the
chains and between them. The ladder has a gap in the excitation
spectrum even in the isotropic coupling limit $J_{\parallel} = J_{\perp}$.The
ground state consists of singlets along the rungs.An excitation
is created by replacing one of the singlets by a triplet and
then letting it propagate. The triplet excitation spectrum
exhibits a gap. Recent inelastic neutron scattering experiments \cite{Eccleston} 
have verified that the S=1/2 AFM vanadyl pyrophosphate
$(VO)_2P_2O_7$ is an accurate realization of the spin-ladder system. 

     A general spin ladder consists of n chains. One example of
such a system is $ Sr_{n-1}Cu_{n+1}O_{2n} $ (n =3,5,7,...) \cite{Gopalan}
which consists of ladders of (n+1)/2 chains with frustrated
"trellis" coupling between the ladders. A ladder with an odd
number of chains has properties similar to that of a single
chain, namely, gapless excitation spectrum and a power-law decay
of the spin-spin correlation function. A ladder with an even
number of chains has a spin gap and an exponential decay of the
spin-spin correlation function. The significant difference between
the properties of odd and even-chain ladders has been verified
in a number of experiments \cite{Dagotto1}. The system 
$La_{4+4n}Cu_{8+2n}O_{14+8n}$ has also a ladder-like structure.Another
compound of interest is $LaCuO_{2.5}$. Initial susceptibility
experiments were interpreted as showing a gap in the excitation spectrum
but subsequent $\mu sr$ and NMR experiments indicate an AFM
transition below $T_N \sim$ 110K \cite{Matsumoto}. Hiroi and Takano
\cite{Hiroi} synthesized the first doped ladder system by
replacing some of the La ions of $ LaCuO_{2.5}$ by Sr. They
observed an insulator to metal transition on doping but no
superconductivity was found down to 5 K. The theoretical studies
on ladders, on the other hand, predict strong superconducting
pairing correlations \cite{Dagotto}.

     The compound $Cu_2(C_5H_{12}N_2)_2Cl_4$ is another example of a
two-chain ladder compound \cite{Hayward}. Magnetic
susceptibility results indicate the presence of weak FM diagonal
interactions in the ladder. The compound $Sr_{14}Cu_{24}O_{41}$
is composed of layers containing two-chain ladders alternating with
layers of $CuO_2$ chains. Experiments have been carried out both on
this system as well as on the system doped with Ca ions: 
 $(Sr_{0.8} Ca_{0.2})_{14} Cu_{24}O_{41}$ \cite{Matsuda, Eccleston1}.
Spin gaps have been seen in the excitation spectra of
both the chains and the ladders.A recent exciting development is
the observation of superconductivity in the ladder compound
$Sr_{0.4}Ca_{13.6}Cu_{24}O_{41.84}$ under a pressure of 3 to 4.5 GPa
\cite{Maekawa}. The superconducting transition temperature is 12
and 9K at 3 and 4.5 GPa, respectively. The discovery of
superconductivity is in conformity with theoretical predictions.

     The compound $CaCuGe_2O_6$ can be described in terms of
isolated dimers (singlets) \cite{Sasago}. There is a finite
energy gap separating the singlet S =0 ground state from the
excited S = 1 triplet. The compound $BaCuSi_2O_6$ is a quasi-2d
AFM with a bilayer structure \cite{Zheludev}. Experiments show the
existence of a spin gap.Dimers predominantly form between the
layers and are weakly-interacting. A recent addition to the list
of AFM systems exhibiting spin gap is the compound $CaV_4O_9$ \cite{Taniguchi}.
The lattice structure of this compound corresponds to the
1/5-depleted square lattice. In this lattice, 1/5 of the original
lattice sites of the square lattice are missing (Fig.1). 

     Lastly, mention should be made of the fact that some of the
high-$T_c$ cuprate systems also exhibit spin gap.A good example is
the bilayer yttrium-barium-copper-oxide compound \cite{Kampf}. 
Detailed discussion of the nature of the spin gap in the cuprate
systems is, however, beyond the scope of this review. 

\section*{III. Exactly-solvable models with spin gap}

     The first example that comes to mind is the celebrated
Majumdar-Ghosh (MG) chain \cite{Majumdar}. The S=1/2 AFM Hamiltonian
includes both n.n. and n.n.n. interactions of strength J and J/2
respectively. 
\begin{eqnarray}
H_{MG}\,=\,J\,\sum_i\,\vec{S_i}\cdot\,\vec{S_{i+1}}
         +\, \frac{J}{2}\,\sum_i\,\vec{S_i}\cdot\,\vec{S_{i+2}}
\end{eqnarray}
The first term in the Hamiltonian is the usual Heisenberg Hamiltonian
for which the ground state energy can be calculated exactly
using the Bethe Ansatz \cite{Bethe}. The ground state is non-degenerate
and disordered. The structure of the ground state is not known
explicitly enough so that calculation of correlation functions is
not possible. For the MG Hamiltonian, the ground state, with periodic
boundary conditions, is doubly degenerate and has a simple structure
\begin{eqnarray*}
\,\,\,\,\,  \phi_1\,&=&\,[\,12\,]\,[\,34\,]......[\,N-1\,N\,] 
\end{eqnarray*}
\begin{eqnarray}
  \phi_2\,&=&\,[\,23\,]\,[\,45\,]......[N\,\,1]
\end{eqnarray}
where [lm] is the singlet ( S=0) with spin configuration $ \frac{1}{\sqrt{2}}
\,\left(\alpha(l)\,\beta(m)\, -\, \beta(l)\, \alpha(m)\right)$, 
$\,\alpha,\,\beta$ being
the spin-up and down spins and l, m are the lattice sites. The
ground state energy $E_g\,=\,-\,\frac{3\,N\,J}{8}$ where N is the number of spins
in the chain. The ground state has total spin {\bf S } = 0. Translational
symmetry is broken in the ground state and the two-spin
correlation function has an exponential decay, i.e., there is no
conventional LRO. 

     The MG chain with simple ground states has motivated a
large number of studies of AFM models with similar ground states
\cite{Shastry,Klein,Caspers}. In all these models the proof
of exact ground state is obtained using the method of `divide and
conquer' \cite{Anderson}. Suppose one is able to construct an
exact eigenstate of a spin Hamiltonian with energy $E_1$. Let $E_g$ and
$ \Psi_g$ be the exact ground state energy and wave
function. Then 
\begin{equation}
               E_g\,\,\leq\,\,E_1
\end{equation}
The Hamiltonian H is divided into sub-Hamiltonians $ H_i$'s for
which the ground state energy $E_{ig}$ can be determined
exactly. Then from the variational theorem
\begin{eqnarray}
E_g\,&=&\,\langle\,\Psi_g\,|\,H\,|\,\Psi_g\rangle 
     \,=\,\sum_{i}\,\langle\,\Psi_g\,|\,H_i\,|\,\Psi_g\,\rangle\,\,\geq\,\sum_{i}\,E_{ig}
\end{eqnarray}
Thus from (9) and (10) one obtains the relation
\begin{equation}
\sum_{i}\,E_{ig}\,\leq\,E_g\,\leq\,E_1
\end{equation}
If one can show that $E_1 = \sum_{i}\, E_{ig}$, then the exact
eigenstate is also the exact ground state. For the MG chain ,the states
$\phi_1$ and $\phi_2$ can be shown to be the exact ground states
by the use of the spin identity
\begin{equation}
{\bf S_n\,\cdot\,(\,S_l\,+\,S_m\,)}\,[l\,m]\,\equiv\,0
\end{equation}
The energy of the exact eigenstate is $E_1$ = $\frac{-\,3\,J\,N}{8}$. The
Hamiltonian $H_{MG}$ is divided into cluster sub-Hamiltonians
$H_i$'s, each $H_i$ describing a triplet of successive spins (123, 
234, 345,...etc.). Each spin in the three-spin cluster (a triangle)
interacts with the other two spins with the strength J/2. The
ground state energy $E_{ig}$ is $\frac{-\,3\,J}{8}$ corresponding to a singlet 
along one of the bonds in the triangle. Since, there are N
sub-Hamiltonians, $ \sum_{i}E_{ig}\,=\,\frac{-3\,J\,N}{8}$ which is equal to
the energy $E_1$ of the exact eigenstate. Thus this state is also
the exact ground state.In adding up the sub-Hamiltonians, the
n.n. bonds are counted twice and so have the strength J in the
total Hamiltonian. The n.n.n. bond which is counted once has
strength $\frac{J}{2}$. 

     Recently, there has been a large number of studies on
frustrated quantum AFMs in 2d which includes both n.n. as well as
further-neighbour interactions \cite{Dagotto2}. The frustrated
models are described as $J_1 - J_2$ and $J_1-J_2-J_3$ models where
$J_1,\,J_2$ and $J_3$ are the strengths of the n.n., diagonal (d) and
n.n.n. exchange interactions. The ground states of these models
in certain parameter regimes are expected to be
spin-disordered. The ground states, though lacking in
conventional LRO, can be characterised by novel order
parameters. Four candidate ground states that have been proposed
are \cite{Bose} chiral, twisted, strip or collinear and
columnar dimer (CD) states. In the fourth type of state, the
ground state consists of dimers (singlets) arranged in
columns. For the square lattice, four such states are
possible. Bose and Mitra \cite{Bose1} have constructed the S =1/2 AFM
$J_1 - J_2 - J_3 - J_4 - J_5$ model on the square lattice for
which the CD states are the exact eigenstates when
\begin{equation}
  J_1: J_2: J_3: J_4: J_5 = 1 : 1: \frac{1}{2} :\frac{1}{2} : \frac{1}{4}
\end{equation}
The proof of exact eigenstate can be obtained using the
spin-identity (12). The CD states are presumably also the exact
ground states. The `divide and conquer' proof, based on
three-spin sub-Hamiltonians, works only for the case when $J_1\,=\,7J$
and all the other interaction strengths are in the ratio given
in (13). Also, only one of the CD states for which the dimer
bonds have the strength 7J is the exact ground state. Mean-field
theory based on the bond-operator formalism shows \cite{Bhaumik}
that the CD states are the ground states when the ratio of
interaction strengths satisfies (13). 

     Bose and Gayen\cite{Bose2} have constructed a spin-ladder
model (Fig.2) which includes diagonal interactions besides n.n.
intra-chain and rung exchange interactions. The rung exchange interaction
has strength $J^\prime$ and the other exchange interactions are of
equal strength J. For $J^\prime\,\geq 2J$, the exact ground state
consists of singlets along the rungs. This is in conformity with
the approximate ground state of the usual spin-ladder which does
not have diagonal exchange interactions. Inclusion of these
interactions makes it possible to determine the exact ground
state. An excitation is created in the model ladder system by replacing
one of the singlets by a triplet. The triplet excitation is
localized and separated by an energy gap from the ground
state. The localized triplet excitation is an artefact of the
special interaction strengths assumed in the spin-ladder model. 
Ghosh and Bose \cite{Ghosh} have generalised the two-chain
ladder system of Bose and Gayen to a n-chain ladder system. In
this system, alternate two-chain ladders have diagonal interactions. 
The chain of rungs in the vertical direction has both n.n. as
well as n.n.n. interactions. The n.n.n. interactions have finite
and zero strengths in an alternate manner. The exact ground state
can be determined for n both odd and even. It can also be
rigorously shown that for n odd(even) the excitation spectrum is
gapless (with gap). This is in conformity with the experimental
results on the ladder system $Sr_{n-1}Cu_{n+1}O_{2n}$ mentioned
in Section II. 

     Recently, Bose and Ghosh \cite {Bose3} have considered the 
S =1/2 AFM Heisenberg Hamiltonian on the 1/5-depleted lattice
which describes the compound $CaV_4O_9$ mentioned in Section II. 
They have constructed a model Hamiltonian which includes both
n.n. as well as further-neighbour interactions. For a
particular ratio of interaction strengths, the plaquette resonating-valence-bond 
(PRVB) state has been shown to be the exact ground state. Various
studies \cite{Katoh,Ueda,Starykh} have indicated that the PRVB
state might be the ground state of $CaV_4O_9$. The
experimentally-observed spin gap in $CaV_4O_9$ can be naturally
linked to the PRVB state. The model proposed by Bose and Ghosh
shows that the PRVB state is an exact ground state. The ground
state is unique with total spin S =0 and does not break any
lattice symmetry. The PRVB state consists of a RVB-type spin
configuration in each plaquette (marked `A' in Fig.1) of the 
1/5-depleted lattice. The RVB state is a linear superposition of
two dimer states. In one dimer state, the spin singlets (dimers
or valence bonds) form along the horizontal bonds. In the other
dimer state, the singlets are along the vertical bonds. 

     We have so far been discussing spin models with spin S
=$\frac{1}{2}$. For a spin-1 chain, Affleck et al \cite{Affleck1} proposed a
model, known as the AKLT model, for which the ground state can
be determined exactly. It can further be rigorously shown that
the excitation spectrum has a gap, thus verifying Haldane's
conjecture. The ground state is unique with no broken
translational symmetry in contrast to the MG chain. To construct
the ground state, regard each spin 1 as a pair of
spin-$\frac{1}{2}$'s. Couple all the n.n. spin-$\frac{1}{2}$'s into singlets. This
state does not have S=1 at each site but this can be remedied by
symmetrizing the wave function at each site. In the final state, 
each n.n. bond has no longer spin S = 0 but it does have the
property that there is no S = 2 component. The Hamiltonian 
$H_{AKLT}$ is written as a sum over projections onto spin 2
($P_2$) of successive pairs of spins, i.e., 
\begin{eqnarray}
H_{AKLT}\,&=&\,\sum_{i}\,P_2\,(\,\vec{S_i}\,+ \,\vec{S_{i+1}}\,)\\
\,&=&\,\sum_{i}\,(\frac{1}{2}\,\vec{S_i}\cdot\,\vec{S_{i+1}}\,
 +\,\frac{1}{6}\,(\vec{S_i}\cdot\,\vec{S_{i+1}}\,)^2 \,+\,\frac{1}{3}\,)
\end{eqnarray}
$H_{AKLT}$ acting on the state described above gives the value
zero as there is no S = 2 component for each n.n. pair of spins. 
Since the projection operator has zero or positive eigenvalue,
the state considered is the exact ground state with eigenvalue
zero. Extension of the S=1 AKLT model to general spin S and to
higher dimensional lattices is possible \cite{Affleck2}. 

\section*{IV. Exact results for doped spin ladders}

     The study of doped AFMs has aquired added significance in connection
with high-$T_c$ cuprate systems. The cuprates in their undoped
state are AFMs and Mott insulators. On doping with a few percent
of holes the AFM LRO is rapidly destroyed giving rise to a
spin-disordered state. The study of frustrated AFMs with
spin-disordered state as ground states is of relevance in this
context. The cuprates have a rich phase diagram as the temperature
and dopant concentrations are varied. The holes predominantly
move in a background of antiferromagnetically interacting spins
residing in the copper-oxide planes, a common structural
ingredient of all the cuprate systems. The phase diagram exhibits, 
besides the insulating phase, metallic as well as
superconducting (SC) phases. In the SC phase, charge transport
occurs through the motion of bound pairs of holes. A large number
of studies exists to understand the dynamics of holes in an
antiferromagnetically interacting spin background. As mentioned
in Section III, Bose and Gayen \cite{Bose4} have constructed a
spin-ladder model (Fig.2) for which the exact ground state
consists of singlets along the rungs. The dynamics of a single
hole  and a pair of holes have been studied in this model \cite{Bose2,
Bose4,Gayen} and several exact results have been obtained. In the
following, we give a brief description of these results. For a
general discussion of hole dynamics in the usual spin-ladder model
(ladder model not containing diagonal interactions), one should
refer to \cite {Dagotto1,Troyer}. 

     The doped spin ladder model is described by the t-J
Hamiltonian given by
\begin{eqnarray*}
H_{t-t^{\prime}-J}\,=\,-\,\sum_{i,j,\sigma}\,t_{ij}\,C_{i\sigma}^{\dag}
C_{j\sigma}\,\,+ \,h.c.\,+\,\sum_{<ij>}\,J_{ij}\,\vec{S_i}\cdot\,\vec{S_j}
\end{eqnarray*}
\begin{eqnarray}
 =\,H_t\,+\,H_{t^\prime}\,+\,H_J\quad\quad\quad\quad\quad\quad
\end{eqnarray}
The constraint that no site can be doubly occupied is implied in
the model. The hopping integral $t_{ij}$ has value t for n.n. 
hopping within a chain and also for diagonal transfer between
chains (solid lines in Fig.2). The corresponding spin-spin interactions, 
$J_{ij}$ are of strength J. The spins have magnitude $\frac{1}{2}$. The
hopping integral across vertical links (dotted lines) connecting
two chains has strength $t^\prime$. The corresponding spin-spin
interaction strength $J_{ij}$ is assumed to be 2J though the
exact results derived below hold true also for other interaction
strengths. In the following, we assume t and $t^\prime$ to be positive. 
In the half-filled limit,i.e., in the absence of holes, the $t-t^\prime-J$
Hamiltonian in (16) reduces to $H_J$. The exact ground state $\Psi_g$
of $H_J$ consists of singlets along the vertical bonds with
energy $E_g\,=\,-\,(\frac{3\,J}{2})\,N$, where 2N is the number of sites in the
system. For $J^\prime$ > 2J, the exact ground state is still the
same; however, for $J^\prime$ < 2J, the state, though an exact
eigenstate, may not be the ground state.We now introduce a
single hole into the system.Let $\Psi(m)$ denote a spin
configuration when the single hole is located in the mth column (rung)
of the ladder model. 
\begin{equation}
\Psi(m)\,=\,\frac{1}{\sqrt{2}}\,(\,\Psi_m(p)\,+\,\Psi_m(q)\,)
\end{equation}
In $\Psi_m(p)$ and $\Psi_m(q)$, the hole is located in the top
and bottom rows, respectively, on the mth column. The other site
in the mth column is occupied by an up spin. The spin
configurations on all the other vertical links are the same as
in $\Psi_g$, namely, singlets. The wave function
\begin{equation}
\Psi\, = \,\frac{1}{\sqrt{N}}\,\sum_{m=1}^N\,\,e^{ikm}\,\Psi(m)
\end{equation}     
is an exact eigenfunction of the total t-J Hamiltonian H with eigenvalue
\begin{equation}
  E\,=\,2t\,\cos(k)\,-t^\prime\,-\,\frac{3\,J}{2}\,(N-1)
\end{equation}
Eqn.(17) describes the bonding combination of hole states in a
rung (momentum wave vector $k_y$ = 0). The anti-bonding hole state
($k_y\,= \pi$) can be constructed as
\begin{equation}
\Psi^\prime(m)\,=\,\frac{1}{\sqrt{2}}\, (\Psi_m(q)\, -\, \Psi_m(p)\,)
\end{equation}
When $H_{t-t^\prime-J}$ in (16) operates on $\Psi^{\prime}(m)$, the
hole accompanied by a free spin-$\frac{1}{2}$ moves one lattice constant
leaving behind a triplet excitation in column m. On further
operating with the $t-t^\prime-J$ Hamiltonian, a closed subspace
of states is generated in each of which the triplet excitation
is localized in the mth column and the hole quasi-particle (hole
+ up-spin) moves one lattice constant to the left or right. One
can write down a set of exact eigenvalue equations for the
propagating hole.

     For J = 0, the eigenvalue equations are similar to those for
a single hopping electron in a 1d chain of atoms with the atom
number `zero' being an impurity atom. The other atoms are located
at positions 1, 2, 3,...and -1, -2, -3,.... The electron can hop from
one atom to its nearest neighbours with amplitude t. The site
energy of the impurity atom is different from that of the other atoms. 
The problem has been extensively discussed in the Feynman 
Lectures, vol.{\bf III} \cite{Feynman} and provides physical insight for
our eigenvalue problem. In the case of the ladder model, the
localized triplet excitation is the `impurity' atom, the hole
accompanied by a free spin-$\frac{1}{2}$ constitutes the propagating
object and the singlets along the vertical links are the `other atoms'
of the lattice.

     The eigenvalue equations  can also be solved exactly for
the case $J\,\neq$ zero. The hole quasi-particle (QP) can be in a
scattering state or form bound and anti-bound (localized but
with energy greater than that of the scattering state) states with
the triplet excitation. For $\,0\,\leq\,J/t <\,0.05$, the bound
state energy is less than the lowest energy corresponding to the
bonding band (Eq.(19)). For J/t > 0.05, the bonding band has a
lower energy. Thus there is a localization-to-delocalization
transition. The propagating hole has a QP character with charge +
e and spin S = $\frac{1}{2}$. There is thus no spin-charge separation, a
hallmark of the interacting electron systems in 1d, the
so-called Luttinger liquids (LLs). Strongly correlated systems in
dimension d > 1 have been conjectured to be LLs. The doped spin
ladder, though strongly correlated, is not a LL but has
properties similar to those of the Luther-Emery model with
gapless charge excitations and a spin gap \cite{Troyer}. 

     We next turn to the case of the spin-ladder doped with a
pair of holes. A brief description of the exact results obtained
in \cite{Ueda} is given (for details refer to the original paper). 
The holes are introduced in the ground state of the model in two
different vertical links so that the dimers along the links are
broken. The two free spins from the broken dimers combine to make
the total spin of the system either S = 1, i.e., a triplet, or S
=0, i.e., a singlet. Consider first the case S = 1 with $S_z$ =
+1. The states S = 1, $S_z$ = 0,-1 are degenerate with the $S_z$
= + 1 state. Let the holes be located in the columns denoted by 
$m_1$ and $m_2$  respectively, where $m_1$ < $m_2$. The eigenfunction
$\psi$ of the $t-t^\prime-J$ Hamiltonian is given by
\begin{equation}
\psi\,=\,\sum_{m_1<m_2}\,a(m_1\,m_2)\,\psi\,(m_1\,m_2)
\end{equation}
wherethe basis function $\psi(m_1\,m_2)$ is given by
\begin{eqnarray}
\psi(m_1,m_2)=\frac{1}{2}\big[\cdot\cdot\cdot\big|_{{}_{(m_1-1)}}
\left({}^\uparrow_O+{}^O_\uparrow\right)_{m_1}\big|_{{}_{(m_1+1)}}
\cdot\cdot\cdot\big|\,\left({}^\uparrow_O+{}^O_\uparrow\right)_{m_2}
|\cdot\cdot\cdot\big].
\end{eqnarray}
The solid vertical lines represent singlets, the arrows stand
for up-spins and the open circles denote holes. We have to solve
the eigenvalue problem $H_{t-t^\prime-J}\,\psi\,=\,E\,\psi$. The
state $\psi(m_1\,m_2)$ belongs to a closed subspace of states
within which $H_{t-t^{\prime}-J}$ operates. The subspace does not
contain the state in which the two holes are located in the same
column. This fact reduces the ladder problem basically to a 1d
one so that the exact Bethe Ansatz \cite{Bethe} technique can be
applied. For the general situation of r holes, the eigenfunction
is a linear combination of the${}^N\,C_r$ functions $\psi(m_1,\cdot\cdot\cdot
\,,m_r):$
\begin{eqnarray}
\psi\,=\,\sum_{m}\,a(m_1.m_2,\cdot\cdot\cdot\,,m_r)\,\psi(m_1,m_2,\cdot
\cdot\cdot\,,m_r)
\end{eqnarray}
Each of the numbers $m_1,....,m_r$ runs over the possible values
1 to N subject to the condition $m_1\,<\,m_2\,<....<m_r$. The general
BA for the r-hole state can be written as
\begin{eqnarray}
a(m_1,m_2,\cdot\cdot\cdot\,,m_r)\,=\,\sum_P\,exp\,[\,i(\sum_{l=1}\,
k_{Pl}\,m_l\,+ 1/2\,\sum_{l<n}^{1,\,r}\,\phi_{Pl,Pn}\,)]
\end{eqnarray}
P is any permutation of r numbers 1,2,....,r. Pl is the number
obtained by operating P on l. The wave vectors $k_i$'s (r in
number) are determined by applying the periodic boundary
condition which leads to r equations. The phase shifts are
determined by demanding that the same BA (Eq.(24)) satifies the amplitude
equations when holes occupy n.n. links. This leads to $\frac{r(r-1)}{2}$ 
equations for the same number of phase shifts $\phi's$ ($\phi_{ij}\,=
\,-\phi_{ji}$). Thus there are $\frac{r(r+1)}{2}$ equations in as many
unknowns which can be solved in an appropriate manner. The
eigenvalues finally obtained correspond to scattering states as
well as anti-bound states of holes.

     We next consider the situation when the free spins from the
broken dimers (after the introduction of two holes in the ground
state) form a singlet with total spin S =0. In this case, the
reduction of the ladder problem to a 1d situation is not
possible as the subspace of states now includes the state in
which two holes occupy the same vertical link, leading to the
possibility of exchange of holes. The BA technique, unique to 1d systems, 
can now no longer be applied. The subspace of states includes
states of the type
\begin{eqnarray*}
\phi(m_1,m_2)&=&\frac{1}{2\sqrt{2}}\big[\big|\cdot\cdot\cdot
\big|\left({}^\uparrow_O+{}^O_\uparrow\right)_{m_{1}}\big|\cdot\cdot\cdot
\left({}^\downarrow_O+{}^O_\downarrow\right)_{m_{2}}\big|\cdot\cdot\cdot\big|
-\big|\cdot\cdot\cdot\big|\left({}^\downarrow_O+{}^O_\downarrow
\right)\big|\cdot\cdot\cdot\left({}^\uparrow_O+{}^O_\uparrow\right)
\big|\cdot\cdot\cdot\big|\big]\\
&&\quad\quad\quad\quad\quad\quad\quad\quad\quad\quad\quad\quad\quad\quad\quad\quad\quad\quad\quad\quad\quad\quad\quad (25)
\end{eqnarray*}
and
\begin{eqnarray*}
\phi(m_1,m_2)\,=\,\big|\big|\cdot\cdot\cdot{}^O_{O_{{}_m}}\cdot\cdot\cdot\big|\big|.
\end{eqnarray*}

Exact eigenvalue equations can be written down in this case and
solved exactly and analytically. The solutions include scattering
states of two holes, bound as well as anti-bound states. The
method of solution cannot be extended to the case of more than
two holes. The rigorous demonstration of the binding of two holes
is of significance in the context of superconductivity both in
the ladder as well as high-$T_c$ cuprate systems. The exact
results obtained also show that the spin gap of magnitude $\frac{3\,J}{2}$ 
is reduced on doping with holes (see Eq.(19)). Exact results are
also available \cite{Bose5} for the case when the spin-ladder model
considered here is doped with a single magnetic or non-magnetic impurity. 
The spin-ladder model under discussion yields a set of exact
results for the dynamics of holes, which are in conformity with
the results for conventional spin ladders, based on approximate
calculations or exact diagonalization of small systems\cite{Dagotto1,Troyer}. 
In the calculations described, quantum fluctuations and the
constraint of no-double-occupancy have been explicitly and
exactly taken into account. Thus the results obtained are
characteristic of both strong correlation and quantum
effects. The possibilty of superconductivity in a ladder system
with spin gap has been experimentally realized as mentioned in
Section II. The crucial question which remains to be settled is
whether the spin gap disappears in the SC phase or survives even
for temperature T < $T_c$. Experimentally, a reduction of the
spin gap with increasing dopant concentration has been observed. 

     In summary, we have described in this review varied AFM
systems which exhibit spin gaps.Some exactly-solvable models
with spin gaps have been determined.Exact results for the hole
dynamics in a ladder model with spin gap have been discussed. The
review is by no means complete and should be supplemented by the
references mentioned. In the last few years, new AFM systems with
spin gaps have been discovered. These systems exhibit a variety
of phenomena like the spin-Peierls transition, metal-insulator transition
and superconductivity, a full understanding of which is as yet
lacking. We may expect that in the coming years new AFM systems
with spin gap will be discovered and a lot of insight gained
about such systems.     
\newpage
\section*{Figure Captions}
\begin{description}
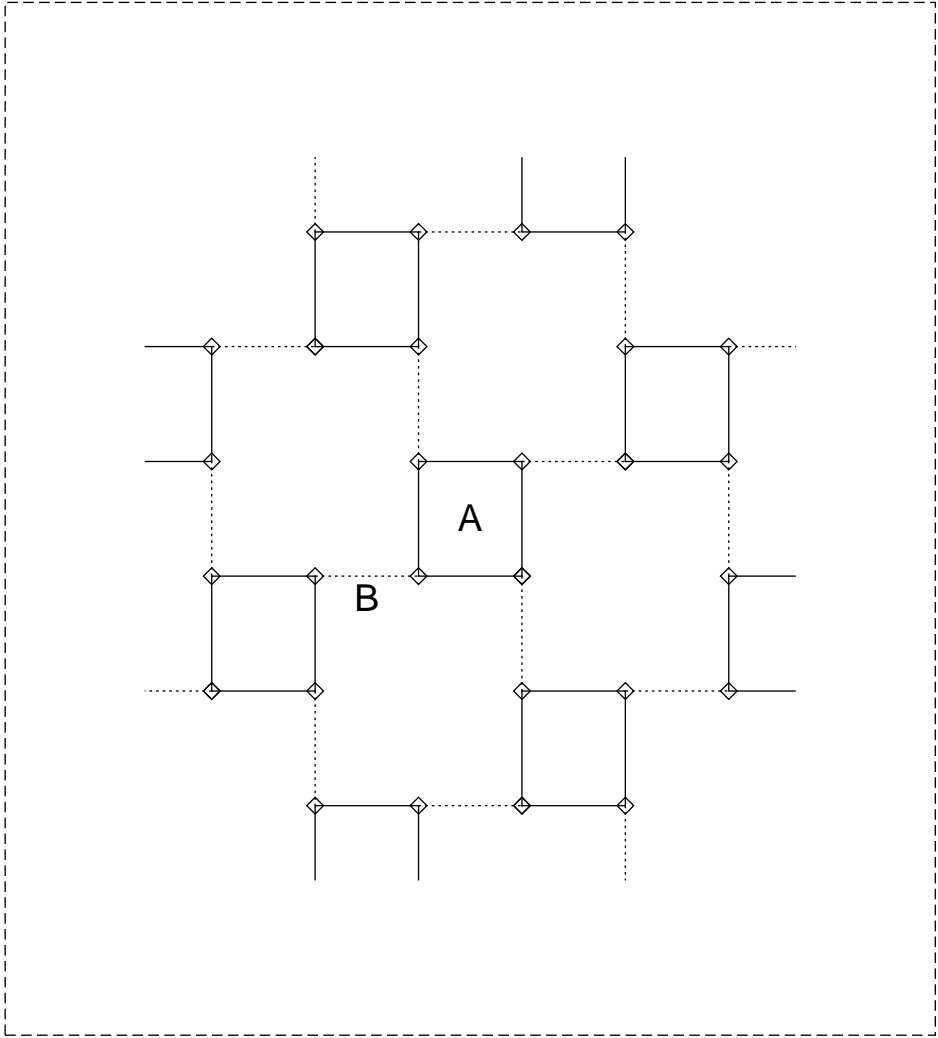
\item[Fig. 1] The 1/5-depleted lattice of $CaV_4O_9$. A and B
represent four-spin plaquettes and dimer bonds connecting
plaquettes, respectively.
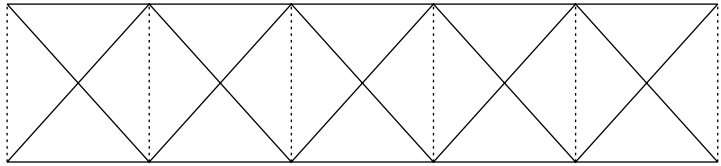
\item[Fig. 2] The spin ladder model described by the
$t-t^\prime-J$ Hamiltonian (Eqn. 16).
\end{description}
  
\end{document}